\documentclass[a4paper,11pt]{article}
\usepackage{amssymb,amsmath,bm}
\usepackage{graphics,graphicx}
\usepackage{array,booktabs}
\usepackage{authblk}
\usepackage{slashed}
\usepackage{subfig}
\usepackage{cite}
\usepackage{xcolor}
\usepackage[margin=2.5cm]{geometry}
\definecolor{Dblue}{rgb}{0.1, 0.1, 0.8}
\definecolor{Lblue}{rgb}{0.22,0.51,0.9}
\definecolor{green}{RGB}{53,170,102}
\definecolor{Green}{rgb}{0.0, 0.5, 0.0}
\definecolor{Bred}{rgb}{0.8, 0.25, 0.33}
\usepackage[linktocpage,
colorlinks = true,
linkcolor = Bred,
urlcolor  = Dblue,
citecolor = green,
anchorcolor = blue]{hyperref}
\usepackage{orcidlink}
\usepackage{epsfig}
\DeclareUnicodeCharacter{2212}{-}

\numberwithin{equation}{section}

\numberwithin{equation}{section}

\title{\textbf{Gravitational wave signatures of  first-order phase transition in two-component dark matter model}}
\author[a]{Seyed Yaser Ayazi\thanks{\href{mailto:syaser.ayazi@semnan.ac.ir}{syaser.ayazi@semnan.ac.ir}}\orcidlink{0000-0002-5994-3316}}
\author[a]{ Mojtaba Hosseini\thanks{\href{mailto:mojtaba\textunderscore hosseini@semnan.ac.ir}{mojtaba\textunderscore hosseini@semnan.ac.ir}}}
\author[a]{Rouzbeh Rouzbehi\thanks{\href{rouzbehrouzbehi@semnan.ac.ir}{rouzbehrouzbehi@semnan.ac.ir}}\orcidlink{0009-0005-7139-5841}}
\affil[a]{Department of Physics, Semnan University, P.O. Box 35131-19111, Semnan, Iran}
\date{\today}

\begin{document}

\baselineskip 0.6 cm
\maketitle

\begin{abstract}
Here, we consider a classically scale-invariant extension of the Standard Model (SM) with two-component dark matter (DM) candidates, including a Dirac spinor and a scalar DM. We probe the parameter space of the model, constrained by relic density and direct detection, and investigate the generation of gravitational waves (GWs) produced by an electroweak first-order phase transition. The analysis demonstrates that there are points in the parameter space, leading to a detectable GW spectrum arising from the first-order phase transition, which is also consistent with the DM relic abundance and direct detection bounds. These GWs could be observed by forthcoming space-based interferometers such as the Big Bang Observer, Decihertz Interferometer Gravitational-wave Observatory, and Ultimate-Decihertz Interferometer Gravitational-wave Observatory.
\end{abstract}

\section{Introduction} \label{sec1}
Impressive progress has been made in the Standard Model (SM) after the discovery of the Higgs boson in 2012 by the ATLAS and CMS experiments~\cite{20121, 201230}, providing us with a deep insight into the building blocks of the Universe. However, there are still unanswered questions, including the matter-antimatter asymmetry, the hierarchy problem, and the nature of dark matter (DM). There is substantial evidence from various cosmological and astrophysical observations indicating the existence of DM~\cite{RevModPhys.90.045002}, which accounts for $\sim$ 27\% of the energy content of the Universe. Weakly interacting massive particles following the freeze-out scenario have been the most popular DM candidates for decades~\cite{10.21468/SciPostPhysLectNotes.71}. Given that, no trace of DM has been found in direct detection experiments so far, and considering the strong bounds on these experiments in one-component DM models, multicomponent DM may present a more promising avenue for future research~\cite{Zurek:2008qg, Profumo:2009tb, PhysRevD.86.076015, Biswas:2013nn, Gu:2013iy, PhysRevD.87.116001, PhysRevD.88.015029, Bian:2013wna, Bhattacharya:2013hva, PhysRevD.89.055021, Esch:2014jpa, PhysRevLett.114.051301, Bian:2014cja, PhysRevD.91.095006, DiFranzo:2016uzc, Aoki:2016glu, DuttaBanik:2016jzv, Pandey:2017quk, Borah:2017aa, Herrero-Garcia:2017vrl, Ahmed:2017dbb, PeymanZakeri:2018zaa, Aoki:2018gjf, Chakraborti:2018lso, PhysRevD.99.015038, PhysRevD.99.076008, Herrero-Garcia:2018qnz, YaserAyazi:2018lrv, PhysRevD.100.015019, PhysRevD.100.103502, Biswas:2019ygr, Nanda:2019nqy, Bhattacharya:2019fgs, Yaguna:2019cvp, Belanger:2020hyh, PhysRevD.103.095016, PhysRevD.102.075039, PhysRevD.103.075001, Hernandez-Sanchez:2020aop, PhysRevD.105.115010, Yaguna:2021vhb, DiazSaez:2021pmg, Kim:2022sfg, Ho:2022erb, Bhattacharya:2022qck, Bhattacharya:2022wtr, Bhattacharya:2018cgx, Bhattacharya:2016ysw, WileyDeal:2023trg, BasiBeneito:2022qxd, Khalil:2020syr, Costa:2022lpy, PandaX:2022djq, Das:2022oyx, Yaguna:2021rds, Berman:2020kko, Becker:2019xzq}. Neither the electroweak phase transition~\cite{Kajantie:1995kf, Kajantie:1996mn} nor the QCD phase transition~\cite{Aoki:2006we, Bhattacharya:2014ara} within the SM are strongly first order, therefore, they do not generate the gravitational wave (GW) signal. However, first-order phase transitions can be caused by various extensions of the SM and the DM candidates, leading to the generation of GWs~\cite{Okada:2018xdh, Croon:2018erz, Madge:2018gfl, Hasegawa:2019amx, Goncalves:2023svb, Abe:2023yte, Cline:1996mga, Costa:2022oaa, Khoze:2022nyt, Hosseini:2023qwu, Hosseini:2024ahe, Mohamadnejad:2021tke, Mohamadnejad:2019vzg, YaserAyazi:2019caf, Asadi:2021pwo,Soni:2016yes, Flauger:2017ged, Chao:2017vrq, Beniwal:2017eik, Huang:2017kzu, Bian:2018bxr, Shajiee:2018jdq, PhysRevD.101.035001, PhysRevD.101.055028, Alanne:2020jwx, PhysRevD.103.035012, PhysRevD.104.053004, Chao:2020adk, Hashino:2018zsi, Baldes:2018nel, Fujikura:2018duw, Athron:2019teq, Alves:2019igs, Goncalves:2022wbp, Ellis:2018mja, Alanne:2019bsm, Paul:2019pgt, PhysRevD.108.055010, Addazi:2019dqt, Kannike:2019wsn, Ertas:2021xeh, Wang:2022akn, Arcadi:2023lwc, Kanemura:2023jiw, Bian:2019szo, Abe:2023zja, Baldes:2018emh, Fairbairn:2019xog, Ghosh:2020ipy, PhysRevD.100.055025, Dunsky:2019upk, Dent:2022bcd, Dichtl:2023xqd, Ahmadvand:2021vxs, Ghoshal:2024gai, Bhattacharya:2023kws, Benincasa:2023vyp, Gu:2023fqm, Bringmann:2023iuz, Han:2023olf, Baker:2019ndr, Breitbach:2018ddu, Baker:2017zwx, Morgante:2022zvc, Schwaller:2015tja, Gao:2023djs, Pasechnik:2023hwv}.

In 2015, the LIGO and Virgo collaboration made the first direct detection of GWs caused by binary black hole mergers~\cite{PhysRevLett.116.061102}. While GWs sourced by strong astrophysical objects can be probed by ground-based GW detectors, upcoming space-based GW interferometers, such as the Laser Interferometer Space Antenna~\cite{LISA:2017pwj}, Big Bang Observer (BBO)~\cite{PhysRevD.72.083005}, Decihertz Interferometer Gravitational-wave Observatory (DECIGO)~\cite{PhysRevD.73.064006}, and Ultimate-DECIGO (UDECIGO)~\cite{PhysRevD.83.044011}, will be able to probe these waves in the millihertz to decihertz range. Detecting such stochastic GWs resulting from cosmological phase transitions in the early Universe could supplement ground experiments like the LHC, thereby opening a unique observational window to explore new physics beyond the SM.

As it is known, the hierarchy problem in particle physics refers to the large discrepancy between the weak force scale ($100~\rm GeV$) and the Planck scale ($10^{19}~\rm GeV$). Resolving these issues is a major motivation for theories beyond SM. In the SM, the Higgs mechanism introduces a scalar field that acquires a nonzero vacuum expectation value, breaking the electroweak symmetry and giving mass to gauge bosons and fermions. However, this requires fine-tuning the Higgs mass parameter to be much smaller than the Planck scale. A scale-invariant solution to this problem proposes that both the Planck and weak scales should emerge as quantum effects rather than fundamental scales in the theory\cite{Foot:2007iy}. In the paper, we present a classically scale-invariant extension of the SM where all the particle masses are generated using the  Coleman-Weinberg mechanism\cite{Coleman:1973jx}. We use the freeze-out mechanism to produce dark matter. The model includes three new fields, two scalars, and one fermion. We probe the parameter space of the model according to constraints from relic density and direct detection. DM relic density is reported by Planck collaboration\cite{Planck:2018vyg} and DM-nucleon cross section is constrained by XENONnT experiment results\cite{XENON:2023cxc}. We study the possibility of the electroweak phase transition with respect to the bounded parameter space, where we use the effects of the effective potential of the finite temperature. We probe the parameter space of the model which is consistent with these constraints and leads to a strong first-order electroweak phase transition. Also, the GW signal resulting from this phase transition has been studied in the BBO, DECIGO, and UDECIGO detectors.

The paper is organized as follows. The two-component DM model is
developed in Sec.~\ref{Model}.  Then the thermal relic density via freeze-out mechanism is calculated in Sec.~\ref{RD}. DM-nucleon cross section is discussed in Sec.~\ref{DD}. Electroweak phase transition and gravitational wave signals are studied in Sec.~\ref{GW}. In Sec.~\ref{Results}, we present the results.  Finally, our conclusion comes
in Sec.~\ref{Conclusion}.

\section{The Model}\label{Model}
In this paper, we review the model presented in \cite{YaserAyazi:2018lrv}. We consider a scale-invariant extension of SM where the Higgs mass term is absent.
Before electroweak symmetry breaking all fields in the scale-invariant sector of potential are massless but as a result of breaking the symmetry of the scale, these fields gain mass from Coleman-Weinberg mechanism\cite{Coleman:1973jx}. We add three new fields, two scalars, and one spinor in the model and all fields are singlets under SM gauge transformation. Two of these new fields, the scalar $S$ and the spinor $\chi$ are assumed to be odd under a $Z_2$ symmetry.
Under $Z_2$ symmetry new fields transform as follows:
\begin{equation}
\phi \to \phi, S \to -S, \chi \to -\chi .
\end{equation}
 The other scalar ﬁeld, $\phi$, and all SM particles are even under $Z_2$. $Z_2$ symmetry guarantees the stability of the lightest odd particles.

The scalar part of the Lagrangian including the new ﬁelds is
\begin{equation}
{\cal{L}}_{scalar}=\frac{1}{2}\partial_\mu\phi\partial^\mu\phi+\frac{1}{2}\partial_\mu S \partial^\mu S+D_\mu H^\dagger D^\mu H - V(H,\phi,S)
\end{equation}
where the most general scale-invariant potential $V(H, \phi, S)$, which is renormalizable and invariant under gauge and $Z_2$  symmetry, is
\begin{equation}
V(H,\phi,S) = \frac{1}{6} \lambda_{H} (H^{\dagger}H)^{2} + \frac{1}{4!} \lambda_{\phi} \phi^4 + \frac{1}{4!} \lambda_{s}S^4 + \lambda_{\phi H} \phi^{2} H^{\dagger}H +  \lambda_{S H} S^2 H^{\dagger}H + \lambda_{\phi s} \phi^2 S^2 \label{2-3}
\end{equation}
where $H$, $\phi$, and $S$ are the doublet Higgs, the scalon, and DM scalars, respectively.
The scale-invariant Lagrangian of the new spinor field and its Yukawa interaction are given by
\begin{equation}
{\cal{L}}_{spinor}=\bar{\chi}(i \gamma^\mu \partial_\mu- g \phi)\chi
\end{equation}
In unitary gauge
\begin{equation}
H = \frac{1}{\sqrt{2}} \begin{pmatrix}
0 \\ h_{1} \end{pmatrix}  \label{2-4}
\end{equation}
and potential (\ref{2-3}) becomes
\begin{equation}
V(h_{1},\phi,S) = \frac{1}{4!} \lambda_{H} h_{1}^{4} + \frac{1}{4!} \lambda_{\phi} \phi^4 + \frac{1}{4!} \lambda_{s}S^4 + \frac{1}{2}\lambda_{\phi H} \phi^{2} h_{1}^{2} +  \frac{1}{2}\lambda_{S H} S^2  h_{1}^{2} + \lambda_{\phi s} \phi^2 S^2 .
\end{equation}
 The Higgs field after spontaneous symmetry breaking is given by
\begin{equation}
H = \frac{1}{\sqrt{2}} \begin{pmatrix}
0 \\\nu_1 + h_{1} \end{pmatrix},
\end{equation}
where $\nu_1 = 246~\rm GeV$. The vacuum expectation value of the field $\phi$ is given by
\begin{equation}
\phi= \nu_2 + h_{2}.
\end{equation}
Notice that $h_1$ and $h_2$ mix with each other and can be rewritten by the mass eigenstates $H_1$ and $H_2$ as
\begin{equation}
\begin{pmatrix}
H_{1}\\H_{2}\end{pmatrix}
=\begin{pmatrix} cos \alpha~~~  -sin \alpha \\sin \alpha  ~~~~~cos \alpha
\end{pmatrix}\begin{pmatrix}
h_1 \\  h_{2}
\end{pmatrix},
\end{equation}
where $ H_{2} $ is along the flat direction, thus $ M_{H_{2}} = 0 $, and $ H_{1} $ is perpendicular to the flat direction which we identify as the SM-like Higgs observed at the LHC with $ M_{H_{1}} = 125 $ GeV. Along the flat direction, the one-loop effective potential has the general form \cite{Gildener:1976ih}
\begin{equation}
V_{T=0}^{1-loop} = a H_{2}^{4} + b H_{2}^{4} \, \log \frac{H_{2}^{2}}{\Lambda^{2}}  , \label{2-13}
\end{equation}
where  $ a $ and $ b $ are the dimensionless constants that given by
\begin{align}
& a = \frac{1}{64 \pi^2 \nu^4}\left(\sum_{bosons}n_j M^4_j (\log \frac{M^2_j}{\nu^2}-\frac{3}{2})-\sum_{fermions} n_j M^4_j (\log \frac{M^2_j}{\nu^2}-\frac{3}{2})\right)  , \nonumber \\
& b = \frac{1}{64 \pi^2 \nu^4}\left(\sum_{bosons}n_j M^4_j -\sum_{fermions} n_j M^4_j\right) , \label{2-14}
\end{align}
and $\Lambda$ is the renormalization group scale. In (\ref{2-14}), $ M_{j} $ and $ n_{j} $ are, the tree-level mass and the internal degrees of freedom of the particle $ j $. By minimizing the relation (\ref{2-13}) and rewriting in terms of the one-loop vacuum expectation value $\nu$, we have
\begin{equation}
V_{T=0}^{1-loop} = b H_{2}^{4} \, \left( \log \frac{H_{2}^{2}}{\nu^{2}} - \frac{1}{2} \right) ,\label{2-16}
\end{equation}
 where $\nu^2=\nu_1^2 + \nu_2^2$. Since in tree level, $M_{H_{2}}=0$,  and the elastic scattering cross section of DM off nuclei becomes severely large, the model is immediately excluded by direct detection experiments. However, at the one-loop level,
radiation corrections give a mass to the massless eigenstate $H_2$\cite{Gildener:1976ih,Ghorbani:2015xvz}:
\begin{equation}
M_{H_{2}}^{2} = \frac{d^2 V_{T=0}^{1-loop}}{d H_{2}^{2}} \bigg\rvert_{\nu}  = -\frac{\lambda_{\phi H}}{16 \pi^2 M_{H_1}^2}(M_{H_1}^4 + M_S^4 + 6 M_W^4 + 3 M_Z^4 - 4 M_\chi^4 - 12 M_t^4).
\end{equation}

After the symmetry breaking, we have the following constraints:
\begin{align}
& \nu_{2} =  \frac{M_{\chi}}{g} , &\nonumber
&sin\alpha = \frac{\frac{\nu_1}{\nu_2}}{\sqrt{1 + (\frac{\nu_1}{\nu_2})^2}},& \nonumber \\
&M_{H_2}= 0,&\nonumber
& \lambda_{H} =  \frac{3 M_{H_{1}}^{2}}{ \nu_{1}^{2}} cos^{2} \alpha, & \nonumber  \\
& \lambda_{\phi} =  \frac{3 M_{H_{1}}^{2}}{ \nu_{2}^{2}} sin^{2} \alpha,  &\nonumber
& \lambda_{\phi H} =  - \frac{ M_{H_{1}}^{2}}{2 \nu_{1} \nu_{2} } sin \alpha  cos \alpha, &\nonumber\\
& \lambda_{SH} = \frac{M_S^2 - 2 \lambda_{\phi s} \nu_2^2}{v_1^2} \label{2-10}
\end{align}
where $M_S$ and $M_\chi$ are the masses of scalar and spinor DM after symmetry breaking, respectively.

According to Eq.~(\ref{2-10}), the model introduces only ﬁve free parameters, $\lambda_s , \lambda_{\phi s} , M_S , M_\chi , g$. In addition, the quartic coupling $\lambda_s$ is irrelevant to the DM relic density. Therefore, the remaining free parameters are
$\lambda_{\phi s} , M_S , M_\chi , g$.

\section{Relic density}\label{RD}
The evolution of the number density of DM particles with time is governed by the Boltzmann equation. The coupled Boltzmann equations for fermion ${\chi}$ and scalar $S$ DM are given by
\begin{equation} \label{44}
\frac{dn_\chi}{dt}+3Hn_\chi = -\sum_{j} \langle \sigma_{\chi\chi \rightarrow jj}\upsilon\rangle (n_\chi ^2 -n_{\chi,eq} ^2)-\langle \sigma_{\chi\chi \rightarrow{S}{S} }\upsilon\rangle (n_\chi ^2-n_{\chi,eq} ^2 \frac{n_{S} ^2}{n_{{S},eq} ^2}) ,
\end{equation}
\begin{equation} \label{45}
\frac{dn_{S}}{dt}+3Hn_{S} = -\sum_{j} \langle \sigma_{{S}{S} \rightarrow jj}\upsilon\rangle (n_{S} ^2 -n_{{S},eq} ^2)-\langle \sigma_{{S}{S} \rightarrow \chi\chi }\upsilon\rangle (n_{S} ^2-n_{{S},eq} ^2 \frac{n_\chi ^2}{n_{\chi,eq} ^2}) ,
\end{equation}
where $j$ runs over SM massive particles, $H_1$ and $H_2$. In $\langle \sigma_{ab \rightarrow cd}\upsilon\rangle$ all annihilations are taken into account except $\langle \sigma_{\chi S} \rightarrow_ {\chi S}\upsilon\rangle$, which does not affect the number density. By using $x=M/T$ and $Y=n/s$, where $T$ and $s$ are the photon temperature and the entropy density, respectively, one can rewrite the Boltzmann equations in terms of  $Y = n/s$:
\begin{equation} \label{46}
\frac{dY_\chi}{dx}=-\sqrt{\frac{45}{\pi}} M_{pl}  g_* ^{1/2} \frac{M}{x^2} [\sum_{j} \langle \sigma_{\chi\chi \rightarrow jj}\upsilon\rangle (Y_\chi ^2 -Y_{\chi,eq} ^2)+\langle \sigma_{\chi\chi \rightarrow{S}{S} }\upsilon\rangle (Y_\chi ^2-Y_{\chi,eq} ^2 \frac{Y_{S} ^2}{Y_{{S},eq} ^2})],
\end{equation}
\begin{equation} \label{47}
\frac{dY_{S}}{dx}=-\sqrt{\frac{45}{\pi}} M_{pl}  g_* ^{1/2} \frac{M}{x^2}[\sum_{j} \langle \sigma_{{S}{S} \rightarrow jj}\upsilon\rangle (Y_{S} ^2 -Y_{{S},eq} ^2)+\langle \sigma_{{S}{S} \rightarrow \chi\chi }\upsilon\rangle (Y_{S} ^2-Y_{{S},eq} ^2 \frac{Y_\chi ^2}{Y_{\chi,eq} ^2})],
\end{equation}
where  $M_{pl}$ is the Planck mass and $g_* ^{1/2}$ is the effective numbers parameter. As can be seen from the above equations, there are new terms in the Boltzmann equations that describe the conversion of two DM particles into each other. These two cross sections are also described by the same matrix
element, so we expect $\langle \sigma_{\chi\chi \rightarrow{S}{S} }\upsilon\rangle$  and $\langle \sigma_{{S}{S} \rightarrow \chi\chi }\upsilon\rangle$ are not independent and their relation is
\begin{equation}
Y_{\chi,eq} ^2 \langle \sigma_{\chi\chi \rightarrow{S}{S} }\upsilon\rangle=Y_{{S},eq} ^2 \langle \sigma_{{S}{S} \rightarrow \chi\chi }\upsilon\rangle .
\end{equation}
We know that the conversion of the heavier particle into the lighter one is relevant, thus the contribution of $\chi$ in the relic density is dominant. The relic density for any DM candidate associated with the $Y$ at the present temperature is given by the following relation:
\begin{equation}
\Omega_{{\chi},S}h^2 = 2.755\times 10^8 \frac{M_{{\chi},S}}{\rm GeV} Y_{{\chi},S}(T_0)
\end{equation}
where $h$ is the Hubble expansion rate at present time in units of 100 (km/s)/Mpc. We implemented the model in the micrOMEGAs package\cite{Belanger:2014vza} to numerically solve the coupled Boltzmann differential equations. According to the data from the Planck collaboration\cite{Planck:2018vyg}, the DM
constraint in this model reads
\begin{equation}
\Omega_{DM} h^{2} = \Omega_{\chi} h^{2}+ \Omega_{{S}} h^{2} = 0.120 \pm 0.001.
\end{equation}
We also define the fraction of the DM density of each component by
\begin{equation}
\xi_\chi= \frac{\Omega_\chi}{\Omega_{DM}}, ~~~~~ \xi_{S}= \frac{\Omega_{{S}}}{\Omega_{DM}}, ~~~ \xi_\chi+\xi_{S} =1.
\end{equation}

\begin{figure}[h!]
\centering
\includegraphics{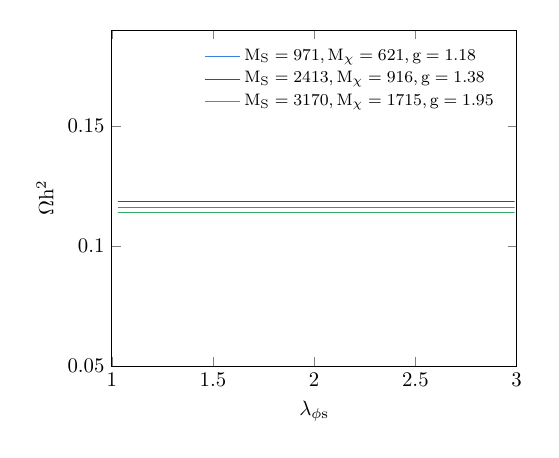}
\caption{The dependency of $\lambda_{\phi s}$ with the total DM relic density.}\label{lambdaphis}
\end{figure}

\begin{figure}[h!]
\centering
\includegraphics[width=0.7\linewidth]{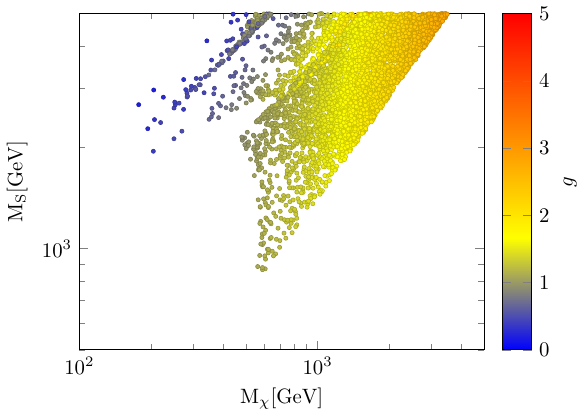}
\caption{The allowed range of parameter space consistent with DM relic density.} \label{Relic}
\end{figure}

In Fig.~\ref{Relic}, the parameter space consistent with DM relic density is obtained. Contrary to the review done in \cite{YaserAyazi:2018lrv}, we have extended the model and investigated more parameter space. As can be seen, there is an agreement with the relic density observed for $100< M_{\chi}<3800$ GeV, $900<M_{S}<5000$ GeV and $0<  g <5$. It is necessary to mention two important points here:\\
(1)  As can be seen from Fig.~\ref{lambdaphis} and it has been investigated in\cite{YaserAyazi:2018lrv}, $\lambda_{\phi s}$ has no effect on the relic density, and for this reason, it has been investigated in Fig.~\ref{Relic} with three quantities of the parameter space. Therefore, we have set $\lambda_{\phi s}$ to be 0.5 throughout the paper.\\
(2)  The mass of the scalar is always greater than that of the fermion, and for this reason(due to conversion $SS \rightarrow \chi\chi$), the fermion($\chi$) occupies a larger share of the observed relic density.

\section{Direct detection}\label{DD}
In this section, we investigate constraints on the parameters space of the model which are imposed by searching for scattering of DM nuclei. The spin-independent direct detection cross sections of $\chi$ and ${S}$ are determined by $H_1$ and $H_2$ exchanged
diagrams\cite{YaserAyazi:2018lrv}:
\begin{align}
& \sigma_{DM-N} ^S =\xi_{S} \frac{\mu_{S} ^2}{4\pi M_{H_1} ^4 M_{H_2} ^4 M_{S} ^2}[\frac{M_{S} ^2 -2\lambda_{\phi s}M_{\chi} ^2 /g^2}{\nu_1}
(\frac{M_{H_2} ^2}{1+(\nu_1 g /M_{\chi})^2 }+\frac{M_{H_1} ^2 g^2 \nu_1^2 }{g^2 \nu_1^2+M_{\chi} ^2 }) \nonumber \\
&+\frac{2 \nu_1 \lambda_{\phi s}}{1+(\nu_1 g /M_{\chi})^2}
(M_{H_1} ^2 - M_{H_2} ^2)]^2 f_N ^2 ,
\end{align}
\begin{equation}\label{direct formula}
\sigma_{DM-N} ^{\chi} = \xi_{\chi} \frac{ g ^3 \nu_1 }{\pi M_{\chi} (1+(\nu_1 g/M_{\chi})^2 )} \mu_{\chi} ^2 (\frac{1}{M_{H_1} ^2}-\frac{1}{M_{H_2} ^2})^2 f_N ^2  ,
\end{equation}
where
\begin{equation}
\mu_{S} = M_N M_S / (M_N + M_S) , ~~~~~ \mu_{\chi}= M_N M_{\chi} / (M_N + M_{\chi}).
\end{equation}
$M_N$ is the nucleon mass and $f_N\simeq 0.3$ parametrizes the Higgs-nucleon coupling.

\begin{figure}[h!]
\centering
	\subfloat[]{\includegraphics[width=0.48\linewidth]{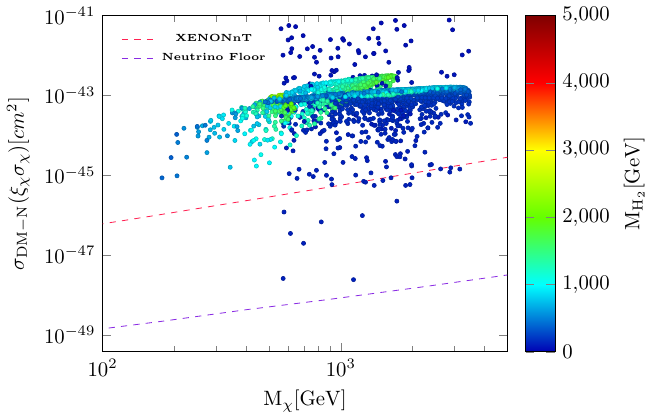}}
	\subfloat[]{\includegraphics[width=0.48\linewidth]{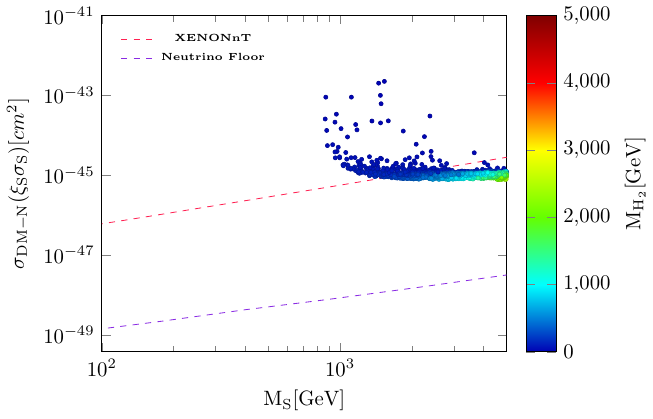}}
\caption{The allowed range of parameter space consistent with DM relic density and direct detection.  In (a) $\xi_S \sigma_S$ vs $M_S$ and in (b) $\xi_{\chi} \sigma_{\chi}$ vs $M_{\chi}$ has shown.} \label{direct detection}

\end{figure}

We use the XENONnT\cite{XENON:2023cxc} experiment results to constrain the parameter space of the model. We have also used the neutrino floor as the irreducible background coming from scattering of SM neutrinos on nucleons\cite{Billard:2013qya}. In Fig.~\ref{direct detection}, rescaled DM-nucleon cross sections($\xi_S \sigma_S $ and $\xi_{\chi} \sigma_{\chi} $) are depicted for the parameters that are in agreement with the relic density. What is clear is that, for large masses, practical parameter space is available for both dark matter candidates.
\section{Electroweak phase transition and gravitational wave signal}\label{GW}
\subsection{Finite temperature one-loop potential}
In addition, to consider the one-loop effective potential at zero temperature, we must compute the one-loop effective potential at finite temperature to discuss electroweak phase transition and gravitational wave. The thermal corrections for finite temperature one-loop potential are given by~\cite{Dolan:1973qd}\footnote{For a review, see~\cite{Laine:2016hma, Croon:2023zay, Athron:2023xlk}.}
\begin{equation}
V^{1-loop}_{T\neq0}(H_2,T) = \frac{T^4}{2\pi^2}\left(\sum_{bosons}n_j J_B(\frac{M^2_j(H_2)}{T^2})-\sum_{fermions}n_j J_F(\frac{M^2_j(H_2)}{T^2})\right) ,
\label{fineffV}
\end{equation}
with thermal functions
\begin{equation}
J_{B,F}(x)= \int_{0}^{\infty} dy  y^2  ln \Bigl(1\mp e^{-\sqrt{y^2+x^2}}\Bigl).
\end{equation}
These thermal functions can be expanded in terms of modified Bessel functions of the second kind~\cite{Mohamadnejad:2019vzg}
\begin{align}
& J_B (x)\simeq -\sum_{k=1}^{3} \frac{1}{k^2}x^2 K_2 (kx) , \nonumber \\
& J_F (x)\simeq -\sum_{k=1}^{2} \frac{(-1)^k}{k^2}x^2 K_2 (kx).
\end{align}
We included the daisy diagrams to improve the validity of perturbation theory~\cite{Khoze:2022nyt}, where $V_{daisy}$\footnote{For theoretical uncertainties exhibits in effective potential including daisy resummation, see~\cite{Gould:2019qek, Croon:2020cgk, Gould:2021oba, Niemi:2021qvp, Schicho:2021gca}} is as follows~\cite{Carrington:1991hz, Arnold:1992rz}
\begin{equation}
V_{daisy}=\frac{T}{12 \pi}\sum_{bosons}n_j\left(M^3_j(H_2)-(M^2_j+\Pi_j(T))^\frac{3}{2}\right)
\label{daisy}
\end{equation}
 The thermal masses, $\Pi_j(T)$ are given by~\cite{Mohamadnejad:2021tke}
 \begin{align}
 & \Pi_W= \frac{11}{6}g_{SM}^2 T^2 \nonumber,~~~~~~~~~~~~~~~~~~~~~~\Pi_z/\gamma=\frac{11}{6}\begin{pmatrix}g^2_{SM}&0\\ 0& g^{\prime2}_{SM}\end{pmatrix}T^2\nonumber,\\
 &\Pi_S=\frac{1}{24}(\lambda_{SH}+6\lambda_S+\lambda_{\phi S})T^2\nonumber ,\\
 &\Pi_{h/\phi}=\frac{1}{24}\begin{pmatrix}(\frac{9}{2}g^2_{SM}+\frac{3}{2}g^{\prime 2}_{SM})+6\lambda^2_t+\lambda_{\phi H}+\lambda_{H S}+6\lambda_H & 0\\ 0 &6g^2+\lambda_{\phi H}+6\lambda_\phi+\lambda_{\phi S}\end{pmatrix}T^2.
 \label{T_corrections}
\end{align}
Finally, the full effective one-loop potential containing (\ref{2-16}), (\ref{fineffV}), and (\ref{daisy}) is given by
\begin{equation}
V_{\rm eff}(H_2 ,T)= V^{1-loop}_{T=0}(H_2) +  V^{1-loop}_{T\neq0}(H_2,T)+V_{daisy}(H_2 ,T).
\label{fullV}
\end{equation}
In order to get $V_{\rm eff}(0, T)= 0$ at all temperatures, we make the following substitution:

\begin{equation}
V_{\rm eff}(H_2 ,T) \rightarrow V_{\rm eff}(H_2 ,T)-V_{\rm eff}(0, T).
\end{equation}
By having the potential~(\ref{fullV}) now we are ready to study the phase transition and GWs associated with that.
\subsection{First-order phase transition and gravitational waves}
Due to the recent advancements in cosmological instruments, the imprints of GWs associated with the first-order phase transitions in the early universe could be detected in the near future. A primary example that encompasses a broad class of phase transitions is electroweak symmetry breaking, which is related to the spontaneous breaking of the gauge symmetry. Electroweak symmetry breaking is a well-studied topic within the context of the SM and can also shed light on the baryon asymmetry of the Universe. Furthermore, the spontaneous breaking of gauge symmetry in the dark sector during the early Universe could result from a first-order phase transition. In this regard, many beyond the SM models predict and study such first-order phase transition and their associated GW signals, including those arising from the dark sector. In the following, we will examine the dynamics of first-order phase transition and identify which parameter points of our model are responsible for such transitions.

The existence of a barrier between the symmetric and broken phases is the first-order phase transition property. When the temperature of the Universe drops below the critical temperature ($T_C$) the electroweak phase transitions can happen. At this temperature, the effective potential~\ref{fullV} has two degenerate minima separated by a high barrier, one in $H_2=0$ and the other in $H_2 = \nu_C \neq 0$:
\begin{align}
& V_{\rm eff}(0 ,T_C)= V_{\rm eff}(\nu_C ,T_C) , \nonumber \\
& \left. \frac{dV_{\rm eff}(H_2 ,T_C)}{d H_2} \right|_{H_2=\nu_C}=0.
\end{align}
One can obtain $\nu_C$ and $T_C$ by solving the above equations. In the model, all independent parameters contribute to the effective potential. However, we find the daisy term is negligible compared to other terms. Therefore $\lambda_{\phi S}$ and $\lambda_S$ are irrelevant and dynamic of the phase transition only depends on $M_S$, $M_\chi$, and $g$~\cite{Mohamadnejad:2021tke}.

At very high temperatures symmetry remains unbroken, and $H_2=0$ represents the true vacuum. As the Universe cools down and the temperature drops below the critical one, an additional vacuum begins to appear. The phase transition from the false vacuum $H_2=0$ to the true vacuum $H_2 \neq 0$ occurs via thermal tunneling at ﬁnite temperature. However, if the barrier is sufficiently high, the tunneling rate may remain very small even at temperatures significantly below the critical temperature. Consequently, it is conventional to deﬁne the nucleation temperature $T_N$, where the corresponding Euclidean action is $S_E= S_3(T_N)/T_N \sim 140$\footnote{It is necessary to treat this condition more carefully at the vacuum-dominated period (see~\cite{Kang:2020jeg}).}~\cite{APREDA2002342}. The theory of such transitions and bubble nucleation was ﬁrst addressed in~\cite{PhysRevD.16.1248, PhysRevD.16.1762}. The function $S_3(T)$ is the three-dimensional Euclidean action for a spherical symmetric bubble given by
\begin{equation}\label{Euclidean action}
S_3(T)= 4\pi \int_{0}^{\infty} dr  r^2 \Biggl(\frac{1}{2} \Bigl(\frac{d H_2}{dr}\Bigl)^2 +V_{\rm eff}(H_2,T)\Biggl),
\end{equation}
where $H_2$ satisﬁes the differential equation that minimizes $S_3$:
\begin{equation}\label{S3}
\frac{d^2 H_2}{dr^2} + \frac{2}{r} \frac{d H_2}{dr}=\frac{dV_{\rm eff}(H_2 ,T)}{d H_2},
\end{equation}
with the boundary conditions:
\begin{equation}
\left. \frac{d H_2}{dr} \right|_{r=0}=0, ~~~~ and ~~~~ H_2 (r\rightarrow \infty )=0.
\end{equation}
We use the publicly available ANYBUBBLE package~\cite{Masoumi:2017trx} to solve the Eq.~(\ref{S3}) and ﬁnd the Euclidean action (\ref{Euclidean action}).

The stochastic GWs associated with the strong first-order phase transition come from three results:\\
(1) Collisions of bubble walls and the resulting shocks in the plasma.\\
(2) Generation of sound waves contributing to the stochastic background after bubble collisions but before the expansion dissipates the kinetic energy in the plasma.\\
(3) Formation of turbulence in the plasma following bubble collisions.\\

These three processes may coexist, and each one contributes to the stochastic GW background:
\begin{equation}\label{GWe}
\Omega_{GW} h^{2} \simeq \Omega_{coll} h^{2}+\Omega_{sw} h^{2}+\Omega_{turb} h^{2}.
\end{equation}

To describe the GW spectrum, we need to define three parameters $\alpha, \beta$, and $\upsilon_\omega$ in addition to the nucleation temperature, $T_N$ that have contributed and controlled the (\ref{GWe}).
 $\alpha$ is the ratio of the free energy density difference between the true and false vacuum and the total energy density given by
\begin{equation}
\alpha= \frac{\Delta \Bigl(V_{\rm eff} -T\frac{\partial V_{\rm eff}}{\partial T}\Bigl)\bigg\vert_{T_N}}{\rho_\ast},
\end{equation}
where $\rho_\ast$ is
\begin{equation}
\rho_\ast= \frac{\pi^2 g_\ast}{30}T_N^4.
\end{equation}
$\beta$ is the inverse time duration of the phase transition given by
 \begin{equation}
\frac{\beta}{H_\ast}= T_N \frac{d}{dT}\Bigl(\frac{S_3 (T)}{T}\Bigl)\bigg\vert_{T_N}.
\end{equation}

Finally, $\upsilon_\omega$ is the velocity of the bubble wall, anticipated to be close to 1 for the strong transitions\cite{Bodeker:2009qy}.

GWs are not generated by isolated spherical bubbles, instead, they arise from collisions between nucleated bubbles. The production of GWs from bubble collisions is given by~\cite{Huber:2008hg}
\begin{equation}
\Omega_{coll}(f) h^{2}=1.67\times 10^{-5} \Bigl(\frac{\beta}{H_\ast}\Bigl)^{-2} \Bigl(\frac{\kappa \alpha}{1+\alpha} \Bigl)^2 \Bigl(\frac{g_\ast}{100} \Bigl)^{-\frac{1}{3}} \Bigl(\frac{0.11 \upsilon_\omega^3}{0.42+\upsilon_\omega^2}\Bigl) S_{coll},
\end{equation}
where $S_{coll}$ parametrizes the spectral shape is given by
\begin{equation}
S_{coll}=\frac{3.8 (f/f_{coll})^{2.8}}{2.8 (f/f_{coll})^{3.8} +1 },
\end{equation}
with
\begin{equation}
f_{coll}= 1.65\times 10^{-5} \Bigl(\frac{0.62}{\upsilon_\omega^2 -0.1\upsilon_\omega +1.8}\Bigl)  \Bigl(\frac{\beta}{H_\ast}\Bigl) \Bigl(\frac{T_N}{100}\Bigl) \Bigl(\frac{g_\ast}{100} \Bigl)^{1/6} Hz.
\end{equation}
The dominant contribution to the GW spectrum is sound waves formed by bubble collisions, given by~\cite{Hindmarsh:2015qta}
\begin{equation}
\Omega_{sw}(f) h^{2}=2.65\times 10^{-6} \Bigl(\frac{\beta}{H_\ast}\Bigl)^{-1} \Bigl(\frac{\kappa_\upsilon \alpha}{1+\alpha} \Bigl)^2 \Bigl(\frac{g_\ast}{100} \Bigl)^{-\frac{1}{3}} \upsilon_\omega S_{sw}.
\end{equation}
The spectral shape of $S_{sw}$ is
\begin{equation}
S_{sw}= (f/f_{sw})^3 \Bigl(\frac{7}{3 (f/f_{sw})^2 +4} \Bigl)^{3.5},
\end{equation}
where
\begin{equation}
f_{sw}= 1.9\times 10^{-5} \frac{1}{\upsilon_\omega} \Bigl(\frac{\beta}{H_\ast}\Bigl) \Bigl(\frac{T_N}{100}\Bigl) \Bigl(\frac{g_\ast}{100} \Bigl)^{1/6} Hz.
\end{equation}
Ultimately, turbulence motion in the plasma resulting from bubble collisions can indeed serve as a source of GWs~\cite{PhysRevD.30.272}. The GW arising from the turbulence is given by~\cite{Caprini:2009yp}
\begin{equation}
\Omega_{turb}(f) h^{2}=3.35\times 10^{-4} \Bigl(\frac{\beta}{H_\ast}\Bigl)^{-1} \Bigl(\frac{\kappa_{turb} \alpha}{1+\alpha} \Bigl)^{3/2} \Bigl(\frac{g_\ast}{100} \Bigl)^{-\frac{1}{3}}  \upsilon_\omega S_{turb},
\end{equation}
where
\begin{equation}\label{Sturb}
S_{turb}= \frac{(f/f_{turb})^3}{(1+8\pi f/h_\ast)(1+f/f_{turb})^{11/3}}
\end{equation}
and
\begin{equation}
f_{turb}= 2.27\times 10^{-5} \frac{1}{\upsilon_\omega} \Bigl(\frac{\beta}{H_\ast}\Bigl) \Bigl(\frac{T_N}{100}\Bigl) \Bigl(\frac{g_\ast}{100} \Bigl)^{1/6} Hz.
\end{equation}
The parameter $h_\ast$ in Eq.~(\ref{Sturb}) is the value of the inverse Hubble time at GW production and accounts for the redshift of the frequency to today,
\begin{equation}
h_\ast= 1.65\times 10^{-5} \Bigl(\frac{T_N}{100}\Bigl) \Bigl(\frac{g_\ast}{100} \Bigl)^{1/6}.
\end{equation}
We choose the following values in computing the GW spectrum based on the suggestion  from numerical simulations~\cite{Caprini:2015zlo, Kamionkowski:1993fg},
\begin{align}
& \kappa= \frac{1}{1+0.715\alpha}(0.715\alpha + \frac{4}{27}\sqrt{\frac{3\alpha}{2}}) , \nonumber \\
& \kappa_\upsilon= \frac{\alpha}{0.73 + 0.083\sqrt{\alpha}+\alpha},~~~~ \kappa_{turb}=0.05\kappa_\upsilon ,
\end{align}
where the parameters $\kappa$, $\kappa_\upsilon$, and $\kappa_{turb}$ denote the fraction of latent heat that is transformed into gradient energy of the Higgs-like field, bulk motion of the fluid, and MHD turbulence, respectively.

\begin{figure}[b!]
\centering
	\subfloat[]{\includegraphics[width=0.48\linewidth]{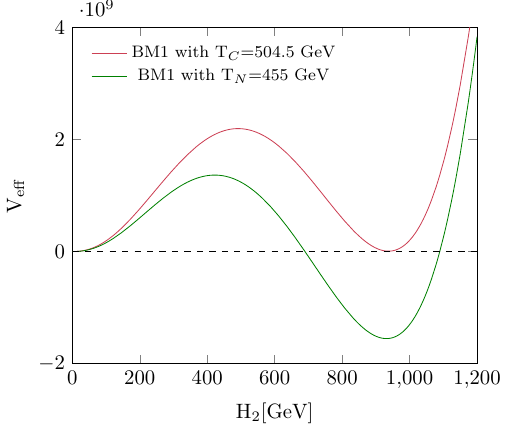}}
	\subfloat[]{\includegraphics[width=0.48\linewidth]{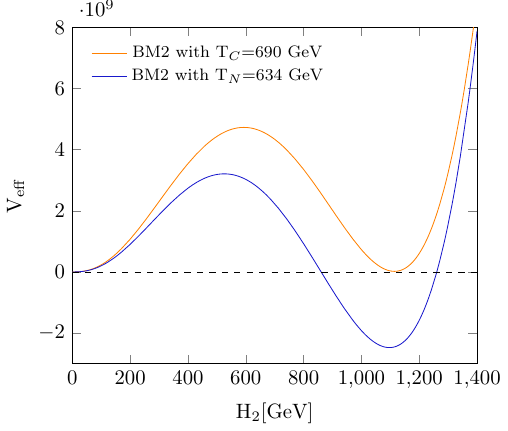}}\\
	\subfloat[]{\includegraphics[width=0.48\linewidth]{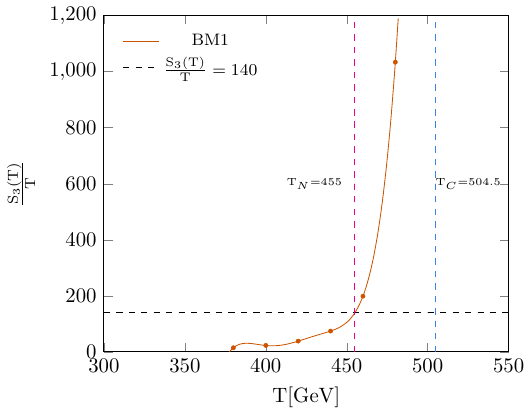}}
	\subfloat[]{\includegraphics[width=0.465\linewidth]{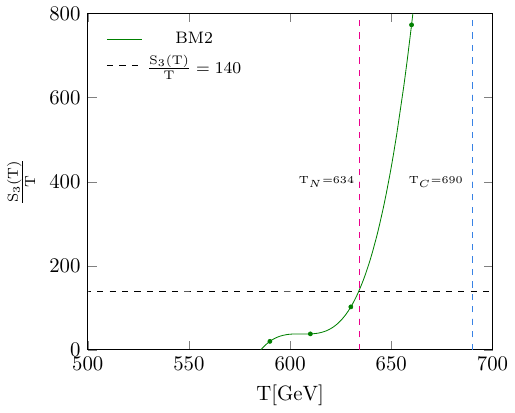}}
		\caption{In (a) and (b) potential behavior are given for critical temperature and nucleation temperature. In (c) and (d), $S_3/T$ changes in terms of temperature are given for all benchmarks.}
	\label{Veffb}
\end{figure}

\section{Results}\label{Results}
\begin{figure}[b!]
\centering
\includegraphics[width=0.7\linewidth]{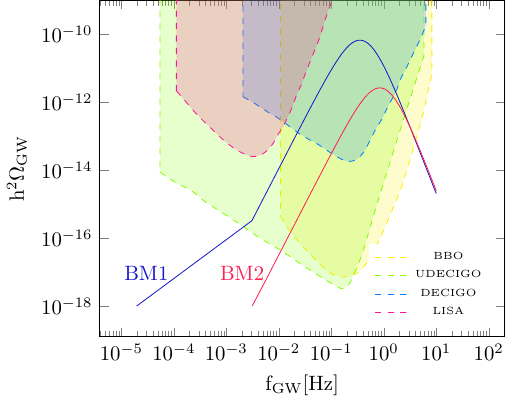}
\caption{GW spectrum for benchmark points of the Table \ref{table}.} \label{GWs}
\end{figure}

To investigate the GWs arising from a first-order electroweak phase transition, we conducted a comprehensive scan of the model's parameter space and selected two benchmark points that satisfy both the relic abundance and direct detection constraints. For the
parameter points in Table \ref{table}, we have compared freeze-out temperature with nucleation
temperature and find out $T_F < T_N$ where $T_F \sim \rm max(M_S, M_\chi )/20$. Therefore, this
issue does not affect our result and the DM properties would not be modified between
$T_F$ and the present day. For each of the selected benchmark points, we computed the parameters $\alpha$, $\frac{\beta}{H_*}$ at the nucleation temperature.  The values obtained for these parameters are presented in Table~\ref{table}. To provide a clear understanding of the phase transition behavior, Fig.~\ref{Veffb} illustrates the effective potential for the chosen benchmark points at both the critical and nucleation temperatures. Additionally, the figure shows the variation of $\frac{S_3}{T}$ in terms of temperature. The primary results of our study, depicted in Fig.~\ref{GWs}, show the predicted GW signals for the selected benchmark points. These signals are within the sensitivity range of upcoming space-based GW detectors such as BBO, DECIGO, and UDECIGO. This indicates that the model's predictions could be tested with the next generation of GW observatories, providing a potential avenue for detecting signatures of new physics related to the electroweak phase transition.

\begin{table}[h!]
\centering
\begin{tabular}{l c c ccrrrrr}
\hline\hline
 $\#$ &$M_S (\rm GeV)$ &$M_{\chi}(\rm GeV)$ &$g$&$M_{H_2}(\rm GeV)$\\
\hline
1&2062&1380&1.748&128.9 \\
2&2215&1501&1.854&129.4\\
\hline\hline
$\#$ &$\Omega_{S} h^{2}$ &$\Omega_{{\chi}} h^{2}$ &$\Omega_{DM} h^2 $&$\xi_S \sigma_S (\rm cm^2)$&$\xi_{{\chi}}\sigma_{{\chi}} (\rm cm^2)$\\
\hline
1&$3.258\times10^{-6}$&$1.186\times10^{-1}$&$1.186\times10^{-1}$&$1.169\times10^{-45}$&$5.563\times10^{-46}$\\
2&$2.834\times10^{-6}$&$1.116\times10^{-1}$&$1.116\times10^{-1}$&$1.246\times10^{-45}$&$7.374\times10^{-46}$\\
\hline\hline
$\#$ &$T_C (\rm GeV)$&$T_N (\rm GeV)$&$\alpha$&$\beta/H_\ast$&$(\Omega_{GW} h^2)_{\rm max}$\\
\hline
1&504.5&455&1.19&4060.06&$1.13\times 10^{-11}$\\
2&690&634&0.36&6976.03&$9.07\times 10^{-12}$\\
\hline
\end{tabular}
\caption{\label{table}Two benchmark points with DM and phase transition parameters.}
\end{table}

\section{Conclusion}\label{Conclusion}
We have considered an extension of the SM  with three new fields: a fermion and two scalars. One of the scalars($S$) together with fermion($\chi$) constitute our dark matter candidates, and the other scalar($\phi$) is considered to be the intermediary between the SM and dark parts. The model is scale invariant and particles are massed through scale symmetry breaking. Therefore, the model can provide a potential solution for the hierarchy problem. The parameter space of the model has been extended to large masses(5 TeV) and a large number of points that agree with the constraints of the relic density and direct detection have been obtained.

We focused our attention on the phase transition dynamics after presenting the model and exploring DM phenomenology. The full finite-temperature effective potential of the model at the one-loop level was obtained to investigate the nature of the electroweak phase transition.  It was demonstrated that the finite-temperature effects induce the first-order conditions of the transition and thereby give rise to a phase transition, which can generate GWs.

After studying the phase transition, we investigated the resulting GWs. We have demonstrated that the model can survive DM relic density and direct detection constraints, while also producing GWs during the first-order electroweak phase transition. We showed GWs for the two benchmark points. These waves can be placed within the observation window of BBO, DECIGO, and UDECIGO. These waves and their investigation in the future can be a hope for new physics.

\bibliography{References}
\bibliographystyle{JHEP}
\end{document}